# Convergent validity of several indicators measuring disruptiveness with milestone assignments to physics papers by experts

Lutz Bornmann* & Alexander Tekles*+

*Division for Science and Innovation Studies

Administrative Headquarters of the Max Planck Society

Hofgartenstr. 8,

80539 Munich, Germany.

Email: bornmann@gv.mpg.de; alexander.tekles@gv.mpg.de

+Ludwig-Maximilians-Universität Munich

Department of Sociology

Konradstr. 6,

80801 Munich, Germany.

**Abstract**

This study focuses on a recently introduced type of indicator measuring disruptiveness in science. Disruptive research diverges from current lines of research by opening up new lines. In the current study, we included the initially proposed indicator of this new type (Wu, Wang, & Evans, 2019) and several variants with $DI_1$: $DI_5$, $DI_{1n}$, $DI_{5n}$, and DEP. Since indicators should measure what they propose to measure, we investigated the convergent validity of the indicators. We used a list of milestone papers, selected and published by editors of *Physical Review Letters*, and investigated whether this human (experts)-based list is related to values of the several disruption indicators variants and – if so – which variants show the highest correlation with expert judgements. We used bivariate statistics, multiple regression models, and (coarsened) exact matching (CEM) to investigate the convergent validity of the indicators. The results show that the indicators correlate differently with the milestone paper assignments by the editors. It is not the initially proposed disruption index that performed best ($DI_1$), but the variant $DI_5$ which has been introduced by Bornmann, Devarakonda, Tekles, and Chacko (2019). In the CEM analysis of this study, the DEP variant – introduced by Bu, Waltman, and Huang (2019) – also showed favorable results.

**Key words**

bibliometrics, convergent validity, disruption index, *Physical Review Letters*

# 1 Introduction

The quantitative study of science using bibliometric indicators (publication and citation data) is widespread today (Fortunato et al., 2018). Research into science is aimed "at advancing our knowledge on the development of science and its communication structure, as well as in relation to social, technological, and socioeconomic aspects" (van Raan, 2019, p. 238). This research mainly began in the 1960s when Eugene Garfield created the Science Citation Index and developed into the broad activity of many scientists with the launch of the Web of Science (WoS, now Clarivate Analytics; Wilsdon et al., 2015). Professional bibliometrics is characterized by the use of advanced bibliometric indicators (Leydesdorff, Wouters, & Bornmann, 2016; van Raan, 2019). These indicators mainly refer to field-normalized indicators such as citation percentiles (Bornmann, 2019). However, the use of these indicators in research evaluation resulted in the critique that "citation-based funding could favor mainstream research and bias decisions against more original projects" (Jappe, Pithan, & Heinze, 2018). Citation counts (and indicators derived from these) might not be able to identify discoveries that are defined by Ziman (1987) as follows: "research results that make a significant change in what we thought we already knew" (p. 98). Highly cited papers might follow short-lasting trends, review the existing literature, or introduce research methods and algorithms (van Raan, 2019).

In recent years, several alternative indicators to citations have been developed measuring novelty of research (Bornmann, Tekles, Zhang, & Ye, 2019). Novelty is related to creativity (novelty is an ingredient of creativity, see Puccio, Mance, & Zacko-Smith, 2013) that involves "the production of high-quality, original, and elegant solutions to complex, novel, ill-defined, or poorly structured problems" (Hemlin, Allwood, Martin, & Mumford, 2013, p. 10). The newly developed novelty indicators are based on the view of creativity as novel recombination of existing elements. For example, Uzzi, Mukherjee, Stringer, and Jones (2013) interpreted more frequent unusual cited references (cited journal) combinations as hints to more novel knowledge (see Lee, Walsh, & Wang, 2015). An overview of the different available approaches in scientometrics to measuring novelty based on unusual combinations of existing elements (e.g. cited references or key words) can be found in Wang, Lee, and Walsh (2018) and Wagner, Whetsell, and Mukherjee (2019).

Based on the fact that impact metrics only measure possible use of new ideas, Funk and Owen-Smith (2017) introduced an indicator (based on patent citations) that is intended to measure newness challenging the existing order. Wu et al. (2019) transferred the idea by Funk and Owen-Smith (2017) to bibliometrics by introducing a disruption index. Azoulay (2019) explains the intuition behind the new indicator as follows: "when the papers that cite a given article also reference a substantial proportion of that article's references, then the article can be seen as consolidating its scientific domain. When the converse is true – that is, when future citations to the article do not also acknowledge the article's own intellectual forebears – the article can be seen as disrupting its domain". A similar index (the so-called dependency indicator, DEP) has been proposed by Bu et al. (2019), who introduced the dependency indicator from a multi-dimensional perspective of impact measurement. According to this perspective, further information from the citing and cited side is considered to measure performance (which is in contrast to the one-dimensional times cited indicator).

The various disruption indicator variants are connected to the distinction by Kuhn (1962) between normal science and scientific revolutions. Kuhn (1962) presents a theory of scientific knowledge: according to Wray (2017), in this theory, "growth of science is not a continuous march closer and closer to the truth. Instead, periods of rapid growth are interrupted by revolutionary changes of theory. These revolutionary changes of theory are disruptive" (p. 66; see also Casadevall & Fang, 2016). In the context of Kuhn's theory of scientific knowledge, Foster, Rzhetsky, and Evans (2015) differentiate between two strategies of conducting research (see also Merton, 1957). The different strategy poles can be denoted as traditional and risky innovative thinking or as convergent and divergent thinking. Divergent thinking increases the chance of conducting breakthrough research. Various papers have been published in scientometrics to date that focus on identifying breakthrough, landmark, or milestone papers (e.g. Chen, 2004; Schneider & Costas, 2016; Thor, Bornmann, Haunschild, & Leydesdorff, in press). An overview of these papers can be found in Winnink, Tijssen, and van Raan (2016).

In this paper, we follow the approach of Bornmann, Devarakonda, et al. (2019) and investigate the convergent validity of the new disruption indicators. We are interested in the question of whether the indicators measure what they intend to measure, namely the disruptiveness of research. The journal *Physical Review Letters* (PRL) published a list of milestone papers published in the journal. The list can be used to validate the indicators: one can expect that milestone papers have higher indicator values than PRL papers not selected for the list. We investigated whether the indicators are able to identify milestone papers and whether there are differences observable between the indicators in this ability.

## 2    Methods

### 2.1    Dataset used

We retrieved the list of milestone papers published in PRL from https://journals.aps.org/prl/50years/milestones. According to a personal communication with Reinhardt B. Schuhmann, the current managing editor of PRL, and Anon (2008), there was an effort to cover all areas of physics research in the list. Many papers have been denoted as milestone papers since they are closely connected to Nobel Prizes in Physics, and occasionally in Chemistry. It cannot be ruled out that bibliometric information played a part in the choice since the selections were made with the benefit of at least ten years of hindsight. However, connections to Nobel Prizes and bibliometrics were not necessary conditions for being a milestone paper. The papers which are listed on the PRL homepage are restricted to papers published until 2002. Since reliable and valid publication and citation data are only available for papers published since 1980 in the in-house database of the Max Planck Society (which is based on the WoS), the focus of this study is on papers on the website published between 1980 and 2002. The dataset of this study consists of all papers published in PRL during this period (see Table 1).

Table 1. Annual number of papers published in *Physical Review Letters* (PRL)

| Publication year | Frequency of all papers | Percent | Frequency of milestone papers |
|---|---|---|---|
| 1980 | 1,072 | 2.39 | 1 |
| 1981 | 983 | 2.19 | 1 |
| 1982 | 1,022 | 2.28 | 4 |
| 1983 | 1,153 | 2.57 | 1 |
| 1984 | 1,247 | 2.78 | 1 |
| 1985 | 1,448 | 3.23 | 1 |
| 1986 | 1,536 | 3.43 | 1 |
| 1987 | 1,461 | 3.26 | 3 |
| 1988 | 1,425 | 3.18 | 3 |
| 1989 | 1,458 | 3.25 | 1 |
| 1990 | 1,641 | 3.66 | 2 |
| 1991 | 1,772 | 3.95 | 2 |
| 1992 | 1,913 | 4.27 | 2 |
| 1993 | 2,091 | 4.67 | 1 |
| 1994 | 1,895 | 4.23 | 1 |
| 1995 | 2,428 | 5.42 | 3 |
| 1996 | 2,684 | 5.99 | 1 |
| 1997 | 2,678 | 5.98 | 1 |
| 1998 | 3,048 | 6.8 | 1 |
| 1999 | 2,854 | 6.37 | 2 |
| 2000 | 3,046 | 6.8 | 4 |
| 2001 | 2,995 | 6.68 | 1 |
| 2002 | 2,962 | 6.61 | 1 |
| Total | 44,812 | 100 | 39 |

We restricted the papers to the document type 'article' in order to standardize the dataset concerning the document type (most of the papers are articles). As Table 1 shows, the dataset of this study consists of 44,812 articles; 39 articles have been classified as milestone papers. Thus the dataset is very unbalanced as to whether or not papers were classified as milestones.

**2.2    Convergent validity**

Validity is one of the most important criteria used to assess bibliometric indicators (Adams, Loach, & Szomszor, 2016; Mutz, 2016; Ruscio, Seaman, D'Oriano, Stremlo, & Mahalchik, 2012; Thelwall, 2017). It is defined as follows: "validity is the extent to which an assessment measures what it claims to measure. It is vital for an assessment to be valid in order for the results to be applied and interpreted accurately" (Panel for Review of Best Practices in Assessment of Research et al., 2012, p. 54). Thus, in this study, we are interested in whether the proposed disruption indicators are able to identify disruptive research. Disruptive research is operationalized by the separation between milestone papers and all other papers published in PRL. The validity examined in this study can be denoted as convergent validity, since we measure the correlation between expert assessments of papers at PRL and bibliometric performance metrics. According to Rowlands (2018), the question in these examinations is "to what extent does a bibliometric exercise exhibit externally convergent and discriminant qualities? In other words, does the indicator satisfy the condition that it is positively associated with the construct that it is supposed to be measuring (i.e. convergent)? The criteria for convergent validity would not be satisfied in a bibliometric experiment that found little or no correlation between, say, peer review grades and citation measures" (Rowlands, 2018).

In recent years, many studies have been published that investigated the correlation between bibliometrics and assessments by peers (e.g. Ahlgren & Waltman, 2014; Baccini & De Nicolao, 2016; Haddawy, Hassan, Asghar, & Amin, 2016; Vieira & Gomes, 2016; Wainer, Eckmann, & Rocha, 2015). An overview of studies in this area can be found in Onodera (2016). The author estimated that the studies mostly resulted in medium correlations (between 0.3 and 0.6). Aksnes, Langfeldt, and Wouters (2019) report some problems of these correlation studies which are also relevant in the context of this study. (1) We cannot exclude the possibility that the PRL peers' assessments are biased – the assignments cannot necessarily be considered as the 'truth'. We tried to consider this point when interpreting the results. (2) Peers frequently use bibliometrics for their assessments and the experts at PRL are no exception (see section 2.1). In order to target this problem, we controlled the number of citations in most statistical analyses.

**2.3   Statistics**

We undertook bivariate analyses in order to investigate the relationship between disruption index variants and citations on the one hand and the milestone assignment to papers on the other. The regression analyses that we also conducted answer the question as to whether this relationship is affected by other variables that are related to the performance of papers (e.g. the number of co-authors). When calculating the regression analyses, we had to decide whether the milestone assignments or the disruption index variants and citations are the dependent variables. Both options can be justified with good arguments. We decided to use the disruption index variants and citations as dependent variables, since these metrics depend on citation decisions across several years. The milestone assessments by the experts are based on the research reported in the papers that can in principle be conducted before performance data emerge. Thus we assume that the performance data follow the research published in papers that can or cannot be classified as milestone. In other words, performance data emerge with the implicit knowledge of significant discoveries that are or are not reported in the papers. Since the decision on the dependent variable is not entirely clear in this study, however, we also report the results of regression analyses (in the appendix) with the milestone paper assignments as the dependent variable.

In the case of disruption index variants and citations as dependent variables, we calculated ordinary least-squares (OLS) regression analyses (Mitchell, 2012). We performed several diagnostics to test the assumptions of these regressions (multicollinearity, heteroscedasticity, and extraordinary cases). In the case of milestone paper assignments as the dependent variable, we applied logistic regression analyses (Long & Freese, 2014). In this study, we are especially interested in the extent to which the milestone paper assignment affects the values of disruption index variants and citations. Since there is a large group of papers in our dataset that are not denoted as milestone papers, we were able to conduct a statistical analysis with a close link to an experimental design (Shadish, Cook, & Campbell, 2002): the matching of observations in treatment and control groups (Austin, 2011; Jann, 2017). The basic idea is to find those papers in the control group (here: the group of papers that are not denoted as milestone papers) that are similar to the milestone papers concerning certain properties (e.g. number of co-authors). It is the goal of the matching approach to receive data with a good balance between treatment and control groups (Iacus, King, & Porro, 2012). Then, the mean difference between both groups can be interpreted as an estimate of the causal effect.

In this study, we applied (coarsened) exact matching (CEM) which requires no assumptions about the underlying data generation process (Jann, 2017). Iacus et al. (2012) explain the method as follows: "The basic idea of CEM is to coarsen each variable by recoding so that substantively indistinguishable values are grouped and assigned the same numerical value (groups may be the same size or different sizes depending on the substance of the problem). Then, the 'exact matching' algorithm is applied to the coarsened data to determine the matches and to prune unmatched units. Finally, the coarsened data are discarded and the original (uncoarsened) values of the matched data are retained". The CEM method is especially indicated for datasets with (very) large control groups (compared to the treatment group).

We used the statistical software Stata to analyze the data (StataCorp., 2017).

**2.4     Description of variables with possible influences on the performance of papers**

We included several variables (e.g. number of co-authors) as control variables in the regression analyses (the variables have been also considered in CEM). In the past, these variables have been identified as factors possibly influencing the citation impact of papers. Since the various disruption indicators that have been considered in this study are based on citations (to a more or less extent), we assume that the indicators also correlate with citations. Since we also know (see above) that citation counts are one criterion for selecting the milestone papers, we controlled citation counts in the regression models. In the following, we describe the variables and report the results of empirical studies dealing with their relationship to citations. More detailed overviews of these studies can be found in Didegah and Thelwall (2013), Onodera and Yoshikane (2014), Tahamtan, Safipour Afshar, and Ahamdzadeh (2016), and Tahamtan and Bornmann (2018a).

Citations appear to be dependent on the number of co-authors. This has been demonstrated by Wesel, Wyatt, and Haaf (2014) for various disciplines. Similar results have been presented by various other studies (e.g. Beaver, 2004; Fok & Franses, 2007; Lawani, 1986; Tregenza, 2002). One reason for the correlation of citations and co-authors might be the higher level of self-citations and network effects (Valderas, 2007). Another variable possibly influencing citations is the number of pages or the length of a paper. Gillmor (1975) reports a corresponding correlation for the *Journal of Atmospheric and Terrestrial Physics*, Stanek (2008) for ecological papers, and Falagas, Zarkali, Karageorgopoulos, Bardakas, and Mavros (2013) for medicine journals. An explanation of these correlations might be that longer papers include more content than shorter papers.

The number of cited references is the third control variable included in this study. Various studies have revealed a positive relationship of this variable with citation counts (e.g.

Ahlgren, Colliander, & Sjögårde, 2018; Fok & Franses, 2007; Peters & van Raan, 1994; Yu & Yu, 2014). Webster, Jonason, and Schember (2009) found that "reference counts explained 19% of the variance in citation counts" (p. 356). An effect of the number of countries involved in a paper and citation counts has been found by Iribarren-Maestro, Lascurain-Sanchez, and Sanz-Casado (2007). Furthermore, the results by Gingras and Khelfaoui (2018) suggest a national citation bias. For certain countries (e.g. the USA) not only higher citation counts can be expected (see also Leydesdorff, Wagner, & Bornmann, 2014), but also collaborations with these countries seem to pay off.

We additionally included the number of years since publishing the paper in the regression models. The number of years should be considered, since not only citation impact (Seglen, 1992) but also disruption index values are time dependent (Bornmann & Tekles, 2019).

## 2.5 Explanations of the citation-based indicators (disruption index variants)

Several indicators have been proposed for measuring the disruptiveness of scientific publications, all following an approach that was first introduced by Funk and Owen-Smith (2017) in the context of patent citation analysis. According to this approach, the disruptiveness of a focal paper is indicated by the extent to which its citing papers do not refer to the cited papers of the focal paper – in this case, the focal paper is the origin for subsequent work without depending on previous work. Wu et al. (2019) were the first to transfer this approach to the context of scientific publications and proposed the indicator $DI_1$ for this purpose, which is defined by the exact same formula that Funk and Owen-Smith (2017) proposed for measuring disruptiveness (see Figure **1**).

For a given focal paper, $DI_1$ is calculated by dividing the difference between the number of publications that cite the focal paper without citing any of its cited references ($N_i$) and the number of publications that cite both the focal paper and at least one of its cited references ($N_j^1$) by the sum of $N_i$, $N_j^1$, and $N_k$ (the number of publications that cite at least one of the focal paper's cited references without citing the focal paper itself). The values of $DI_1$ range from -1 to 1. High (positive) values of $DI_1$ are intended to indicate disruptive research, while small (negative) values are intended to indicate developmental research.

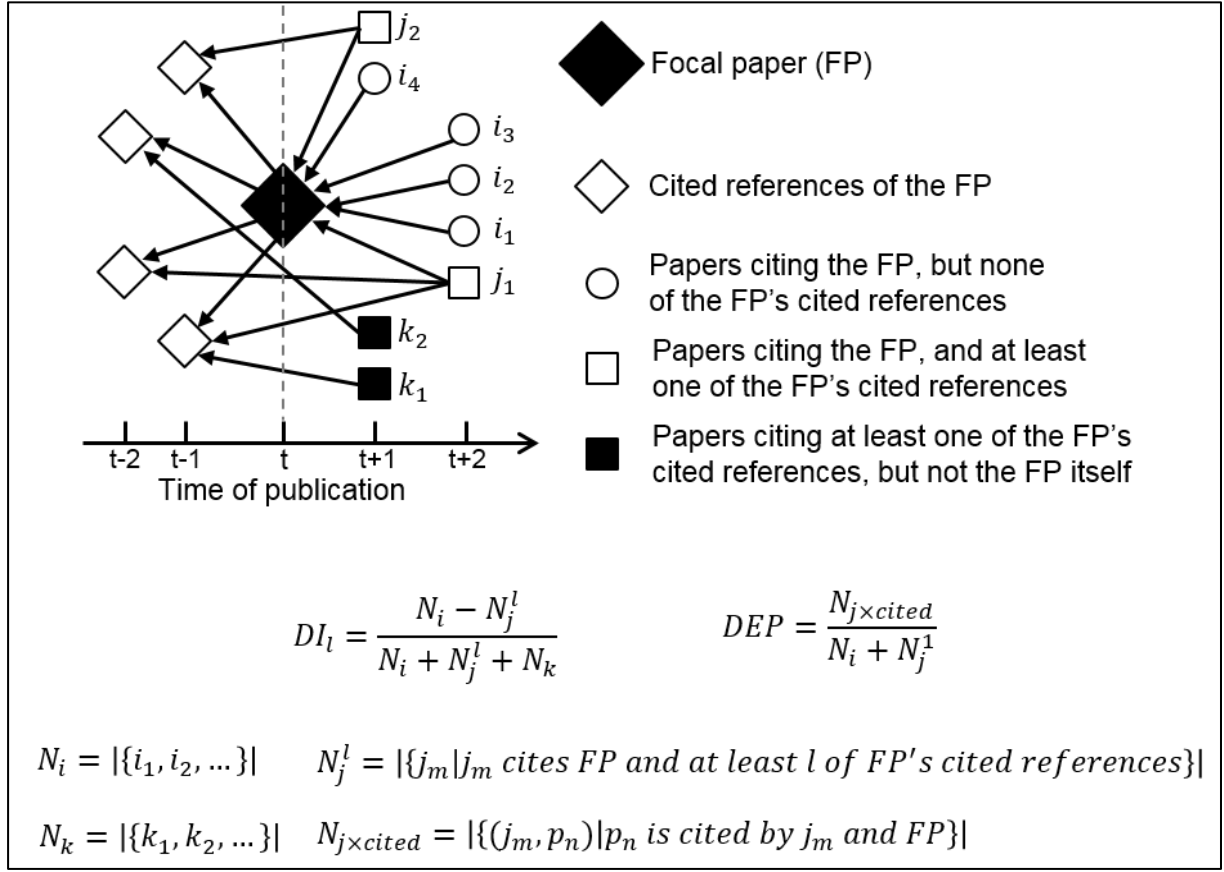

Figure 1. Definitions for disruption indexes $DI_1$ and $DI_5$ as well as dependency indicator (DEP)

Since $DI_1$ has been introduced, several other indicators that follow the same basic approach have been proposed. In the current study, we consider some of these indicators that seem to be promising alternatives according to empirical analyses of various variants (Bornmann, Devarakonda, et al., 2019; Bornmann, Devarakonda, Tekles, & Chacko, 2020). One of these indicators that we include in this study is $DI_5$. In contrast do $DI_1$, $DI_5$ considers how strong the ties between citing and cited papers are for a focal paper. This may mitigate unintended effects of the citation behaviour of a given focal paper. Suppose that a focal paper cites a few highly cited papers, which are very likely to be cited by papers citing the focal paper, even if the focal paper is rather disruptive. In such a situation, the citing papers with only a few citation links to the focal paper's cited references may not be adequate indicators for disruptive research. For the calculation of DI$_5$, only citing papers with at least five citation links to the focal paper's cited references are considered for the term $N_j^5$. In an empirical analyses by Bornmann, Devarakonda, et al. (2019), DI$_5$ produced comparably good results.

According to Bornmann et al. (2020), a modified variant of DI$_5$, the indicator DI$_{5n}$, seems to be able to identify landmark papers in scientometrics. In contrast to DI$_5$, DI$_{5n}$ focuses on the disruptive effect in the focal paper's field. This is achieved by considering the cited references of all papers published in the same journal and the same year as the focal paper for determining $N_j^5$ and $N_k$. In the current study, we consider DI$_{5n}$ as well as DI$_{1n}$ (which is obtained by applying the same modification to DI$_1$) for our empirical analyses.

Another indicator that has been proposed independently of the aforementioned indicators is DEP, introduced by Bu et al. (2019). DEP also follows the basic approach of considering to which extent the citing papers of a focal paper refer to its cited references. Like DI$_5$, DEP considers how strong the ties between the citing and cited papers of a focal paper are. Instead of defining a hard threshold for the minimum number of citation links, DEP counts the number of citation links from papers citing the focal paper to papers cited by the focal paper. The interpretation of DEP differs from the DI$_x$ variants in three points. First, high values of DEP indicate a high dependency of the focal paper on earlier work. Consequently, small values indicate that a focal paper does not depend on earlier work, which can be interpreted as disruptiveness. In order to facilitate the comparison of DEP with the DI$_x$ variants, we inverted DEP by subtracting each value from the maximum value (plus 1).

Second, DEP values range from 0 (which is the smallest possible value DEP can take according to the definition) to an empirical maximum value. There is no theoretical upper bound for DEP. Third, it should be borne in mind that DEP does not consider the citation impact of the focal paper, while the DI$_x$ variants do. Since $N_k$ captures the citation impact of the focal paper's cited references, a focal paper needs to have a relatively high citation impact (compared to its cited references) in order to score high on the DI$_x$ variants. In contrast, for the calculation of DEP, the citing papers of a focal paper's cited references are only considered if they also cite the focal paper.

## 3    Results

In this study, we included four variants of the disruption index and citation counts. We are interested in the question as to how these performance measures correspond to the qualitative milestone paper assessments by the PRL. Figure 2 shows two distributions of the disruptive index variants and citations: (1) the bars are actual distributions, and (2) the smooth, bell-shaped distribution outlines how the data would be distributed if they were normal. We transformed two variables in Figure 2 for the statistical analyses. As mentioned in section 2.5, we inverted DEP by subtracting each value from the maximum value (plus 1). The citation variable is much skewed with a minimum of 0 and a maximum of 74,187 citations (M=93.92, SD=410.18). Lundberg (2007) recommends to calculate logarithmically transformed citation counts [log($x$+1)] to normalize the skewed citation distribution.

The results in Figure 2 confirm the so-called "bibliometric laws" that are defined as follows: "Between the 1920s and 1930s, three milestone studies in the history of the discipline were published, respectively, by Alfred Lotka on the distribution of scientific productivity, by Samuel Bradford on the scattering of papers among disciplinary journals, and by George Zipf on the statistical properties of text corpora. From different starting points and analytic perspectives, the three authors formalized a set of regularities – the 'bibliometric laws' – behind the processes by which a certain number of items (scientific papers, text words) are related to the sources generating them (authors, journals, text corpora). Their common feature is an amazingly steady tendency to the concentration of items on a relatively small stratum of sources" (de Bellis, 2009, p. xxiv).

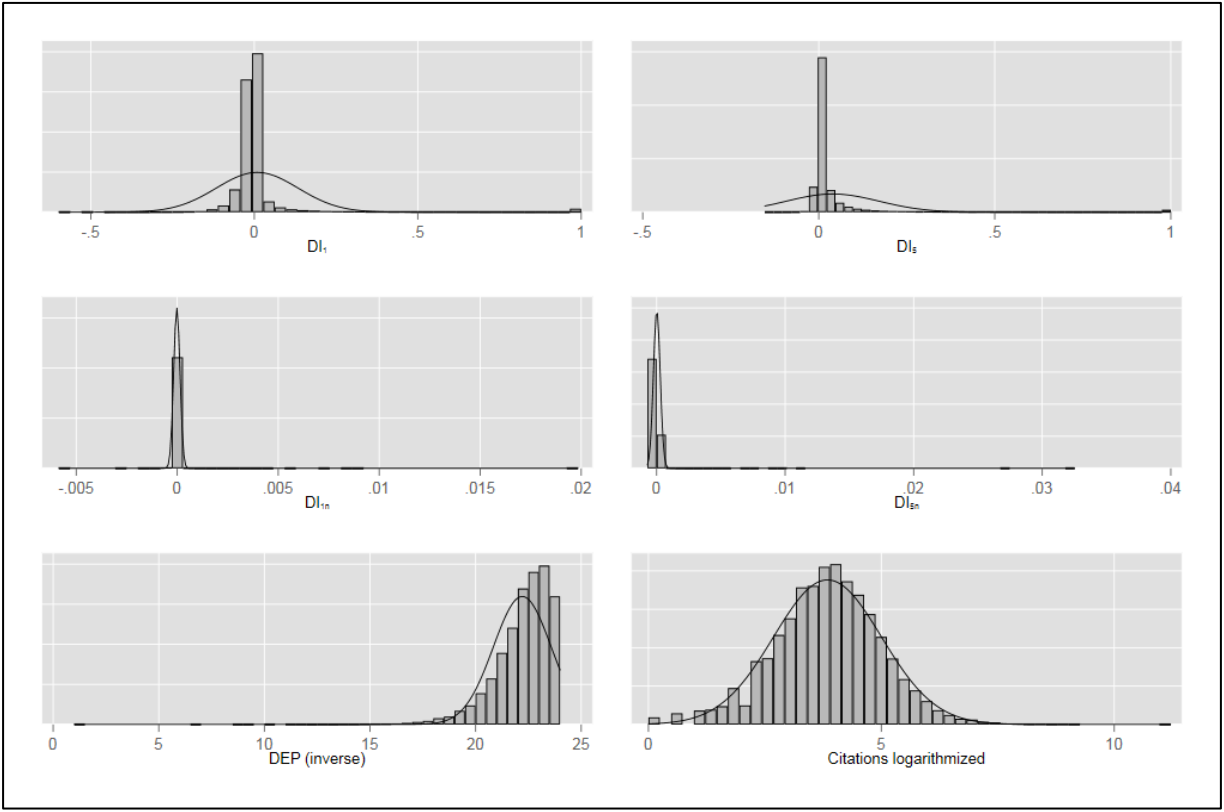

Figure 2. Distributions of disruptive index variants and logarithmized citations

The distributions of both $DI_1$ variants are characterized by many papers with values around zero and only a few papers with very high positive and very small negative values. The distributions of both $DI_5$ variants are similar, but the outliers are concentrated on the positive side. The distribution of the inverse DEP is not concerned by these outliers with high positive or negative values. Instead, the results demonstrate an increasing number of papers that are cited independently from their cited references. In other words, the cited references of the citing papers and the cited references of the focal papers are increasingly (i.e. for more and more papers) independent from each other. The logarithmized citations approximate more or less the estimated normal distribution with the exception of a few very highly cited papers (based on logarithmized citations).

We would like to deal now with the question of how the milestone papers are positioned in the distributions of the various indicators. The results are presented in Figure 3. For every indicator, the medians (light grey dotted line), 90th percentiles (grey dotted line), and 99th percentiles (dark grey dotted line) are shown. Then, the indicator values for the milestone papers are included in every graph. Since up to four papers are available as milestone papers in one year, annual medians have been calculated. As the results in Figure 3 indicate, logarithmized citations performed better than the other (disruption) indicators: most of the milestone papers are above the 99th percentile line. One reason for these results might be that citations were one criteria (among others) for the selection of the milestone papers published in PRL (see section 2.1). A similar favorable result is visible for $DI_{5n}$ where many milestone papers are positioned close to or above the 99th percentile line.

The results in this graph also indicate a time dependency of disruption, with high disruptive values for older milestone papers. It appears that disruptive research emerges after decades. However, the time dependency may also be caused by a time dependency of the citation counts for a focal paper's cited references. For example, if the coverage of the data increases with time, or citation counts generally increase with time, then the term $N_k$ increases with time, resulting in less extreme disruption scores for more recent publications. Therefore, controlling the age of papers is important for interpreting disruption scores (as we did in the regression models of this study).

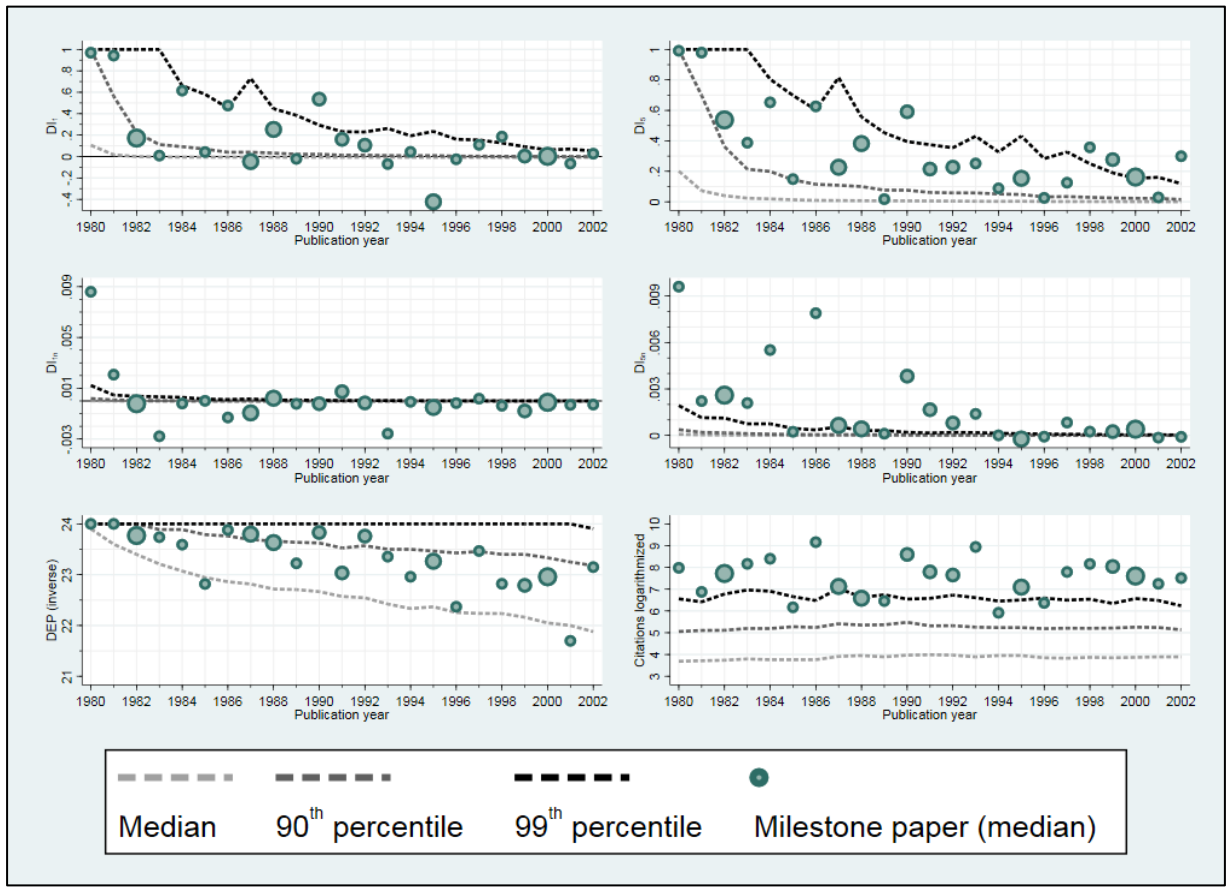

Figure 3. Positions of milestone papers in the distributions of disruptive index variants and logarithmized citations. The graphs show medians, 90[th] percentiles, and 99[th] percentiles of the disruption index variants and logarithmized citations across the publication years 1980 to 2002

Whereas in Figure 3 most of the DEP milestone papers are between the median and 90[th] percentile lines, the $DI_5$ milestone papers are mostly between the 99[th] and 90[th] percentile lines. Most of the milestone papers in the $DI_1$ graph are above the median line. In contrast to $DI_5$ and $DI_{5n}$, $DI_1$ and $DI_{1n}$ are concerned by larger negative values which might denote developmental research (see section 2.5). There are also some negative values for $DI_1$ and $DI_{1n}$ among the milestone papers. This result might suggest that the PRL journal indicated not only disruptive papers as milestone papers, but also research continuing existing research lines. The potential to identify developmental papers seems to be given especially for $DI_{1n}$ where a comparatively large number of papers have negative values.

### 3.1 Regression models

The results visualized in Figure 3 suggest that there is a relationship between disruption in terms of milestone assignments and the disruption indicator values. However, the same results may well be produced even if the disruption index variants are not good indicators for disruptiveness. According to the results, milestone papers generally have a high citation count. However, this does not mean that a highly cited paper is automatically disruptive (or that citation counts are a better indicator for disruptiveness than the disruption indicators). One would also get the same results if only highly cited papers are detected as milestone papers and the disruption index variants only detect highly cited papers to a certain degree instead of disruptive papers.

In order to address this problem, we took a more detailed look at the relationship between milestone paper assignments by the journal and the various indicators considered here. We calculated regression models to investigate this relationship by considering further variables which might have an influence on the performance of papers (see section 2.4). This allows us to assess whether the indicators are able to detect disruptive papers from a set of papers that are similar with regard to these variables. The dependent variables are the various disruption index and citation indicators. The variance of these indicators should be explained by the independent variables (variables possibly influencing paper performance and the binary milestone variable).

We used the power analysis in multiple regression to identify the sample size necessary in order to explain a small proportion of variance in the dependent variables. One needs larger samples to find small effects. This study is based on a very large sample (see section 2.1) and we do not expect (very) large effects. We assume statistically significant independent variables (explaining the dependent variables) on the *p* = 0.05 level and a power of 0.8: the power denotes the chance of finding statistically significant results for the independent variables which are actually related to the dependent variables (see Acock, 2018). With a power of 0.8 and a *p* value of 0.05, one needs around 800 papers to reach an $R^2$ of 0.2 (see Dattalo, 2008). Since this study is based on a much larger sample than 800 papers, we can expect statistically significant results for the independent variables even if the explained variance in the dependent variable is small. In other words, the very large sample size in this study means that very small effects will also be statistically significant.

Before the results of the regression analyses are presented below, we shall elaborate on some assumptions that should be fulfilled in order to trust the conclusions from the models. First, we tested for multicollinearity by using the variance inflator factor for each independent variable. As the results show, the factors are below the critical factor (10) in each case (Sheskin, 2007). Second, we tested for heteroscedasticity by using the Breusch-Pagan and Cook-Weisberg tests for heteroscedasticity (Breusch & Pagan, 1979; Cook & Weisberg, 1983). Since the results were statistically significant for all regression models, we assume that the variances of the residuals are not constant. Heteroscedasticity usually occurs in case of skewed distributions in dependent variables. The results in Figure 2 even revealed these non-normal distributions, and further tests for normality based on skewness and kurtosis confirmed these distributions (Dagostino, Belanger, & Dagostino, 1990). Acock (2018) recommends robust regressions in case of skewed dependent variables (which we followed). In robust regressions, the so-called sandwich estimator is used to estimate the standard errors (Angeles, Cronin, Guilkey, Lance, & Sullivan, 2014; Colin Cameron & Miller, 2015).

In the final test of the regression models, we checked whether there were extraordinary cases (outliers) in the data. We computed Cook's distance index to detect unusual variable constellations in single cases (Sheskin, 2007). The results of the tests are shown in Figure 4. In all models, at least one influential case is detected. Since all models should be performed based on the same publication set, we decided to exclude some cases which are remarkable across the models in the figure: IDs 1511, 23790, 24595, 32492, 33671, and 37444. We performed the models without these papers to check the robustness of the results. The results excluding the IDs can be found in the appendix of this paper. As the results reveal, they did not significantly change when these cases are excluded from the analyses.

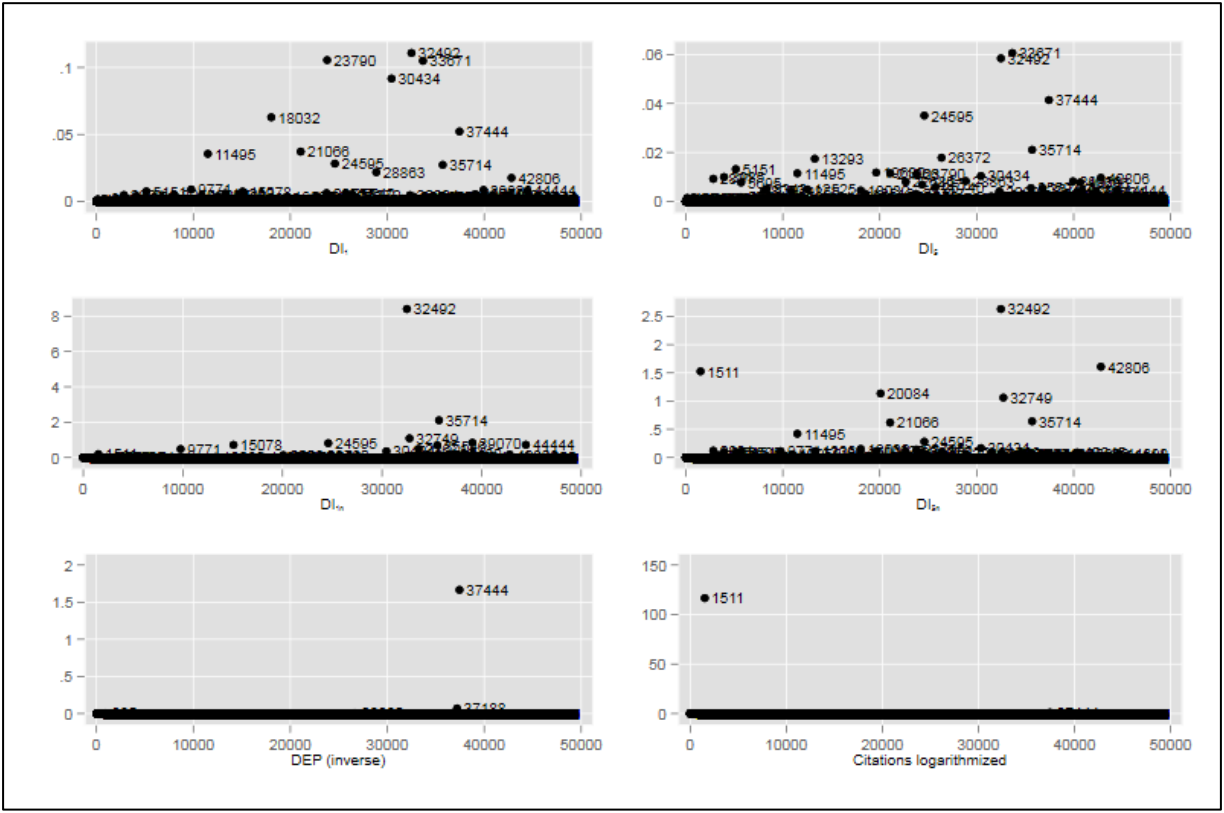

Figure 4. Results of the Cook's distance index for six regression models

The results of the regression models are shown in Table 2. For each disruption index variant and logarithmized citations, two models – with and without other independent variables besides the milestone variable – have been estimated. We are especially interested as to what extent the milestone paper assignment effects the values of disruption index variants and logarithmized citations. The coefficients of this variable are very different in the table depending on the scaling of the dependent variable.

Table 2. Predictors of disruption index variants and logarithmized citations

| Variable | $DI_1$ | $DI_1$ | $DI_5$ | $DI_5$ | $DI_{1n}$ | $DI_{1n}$ | $DI_{5n}$ | $DI_{5n}$ | DEP (inverse) | DEP (inverse) | Citations logarithmized | Citations logarithmized |
|---|---|---|---|---|---|---|---|---|---|---|---|---|
| Milestone letter (=1) | 0.13206* | 0.11606* | 0.29612*** | 0.22827*** | -0.00002 | 0.00002 | 0.00160*** | 0.00148*** | 1.12008*** | 0.24692*** | 3.74091*** | 3.68047*** |
|  | (0.05) | (0.05) | (0.04) | (0.04) | (0.00) | (0.00) | (0.00) | (0.00) | (0.10) | (0.07) | (0.16) | (0.16) |
| SD | 0.12507 | 0.12507 | 0.13170 | 0.13170 | 0.00016 | 0.00016 | 0.00026 | 0.00026 | 1.36918 | 1.36918 | 1.15728 | 1.15728 |
| Number of years |  | 0.00519*** |  | 0.00706*** |  | 0.00000** |  | 0.00000*** |  | 0.07156*** |  | -0.00374*** |
|  |  | (0.00) |  | (0.00) |  | (0.00) |  | (0.00) |  | (0.00) |  | (0.00) |
| Citations logarithmized |  | 0.00101 |  | 0.01429*** |  | -0.00001*** |  | 0.00003*** |  | 0.20242*** |  |  |
|  |  | (0.00) |  | (0.00) |  | (0.00) |  | (0.00) |  | (0.01) |  |  |
| Number of co-authors |  | 0.00001 |  | 0.00005*** |  | 0.00000 |  | -0.00000 |  | 0.00152*** |  | -0.00167** |
|  |  | (0.00) |  | (0.00) |  | (0.00) |  | (0.00) |  | (0.00) |  | (0.00) |
| Number of pages |  | 0.00347** |  | -0.00617*** |  | 0.00000 |  | -0.00001 |  | -0.26398*** |  | 0.23990* |
|  |  | (0.00) |  | (0.00) |  | (0.00) |  | (0.00) |  | (0.04) |  | (0.12) |
| Number of countries |  | -0.00194** |  | -0.00154* |  | -0.00000 |  | -0.00000 |  | -0.03374*** |  | -0.01984** |
|  |  | (0.00) |  | (0.00) |  | (0.00) |  | (0.00) |  | (0.01) |  | (0.01) |
| USA (=1) |  | -0.00359** |  | -0.00438*** |  | -0.00000** |  | -0.00000** |  | -0.01145 |  | 0.24368*** |
|  |  | (0.00) |  | (0.00) |  | (0.00) |  | (0.00) |  | (0.02) |  | (0.01) |
| China (=1) |  | 0.00577** |  | 0.00280 |  | 0.00000* |  | 0.00001** |  | -0.02611 |  | -0.05684 |
|  |  | (0.00) |  | (0.00) |  | (0.00) |  | (0.00) |  | (0.06) |  | (0.05) |
| EU 28 (=1) |  | -0.00097 |  | -0.00009 |  | 0.00000* |  | -0.00000 |  | -0.00207 |  | 0.16220*** |
|  |  | (0.00) |  | (0.00) |  | (0.00) |  | (0.00) |  | (0.02) |  | (0.01) |
| Constant | 0.00858*** | -0.05319*** | 0.03933*** | -0.05473*** | -0.00002*** | 0.00002* | 0.00001*** | -0.00012*** | 22.20830*** | 21.83097*** | 3.84069*** | 2.78272*** |
|  | (0.00) | (0.00) | (0.00) | (0.00) | (0.00) | (0.00) | (0.00) | (0.00) | (0.01) | (0.13) | (0.01) | (0.46) |
| $R^2$ | 0.00097 | 0.06941 | 0.00440 | 0.13074 | 0.00002 | 0.00934 | 0.03235 | 0.06269 | 0.00059 | 0.13670 | 0.00909 | 0.04540 |
| $N$ | 44,809 | 44,809 | 44,809 | 44,809 | 44,812 | 44,812 | 44,812 | 44,812 | 43,849 | 43,849 | 44,812 | 44,812 |

Notes. Robust standard errors in parentheses
* $p < 0.05$, ** $p < 0.01$, *** $p < 0.001$

For example, the (statistically significant) $DI_1$ coefficient for the milestone variable is 0.13, meaning that for a milestone paper, we would expect the $DI_1$ score to be 0.13 units higher. A value of 0.13 corresponds exactly to a standard deviation (SD) of $DI_1$ (the SDs have been added to Table 2). The coefficient is slightly reduced to 0.12 when the other independent variables are included in the model (i.e. holding constant). This reduction can be seen for all models in Table 2 with two exceptions: (1) in the citations models, we have nearly the same coefficients. The reason might be that this is the only model in which citations are not considered as independent variable. (2) The coefficient of $DI_{1n}$ changed from negative to positive (but is close to zero in both cases).

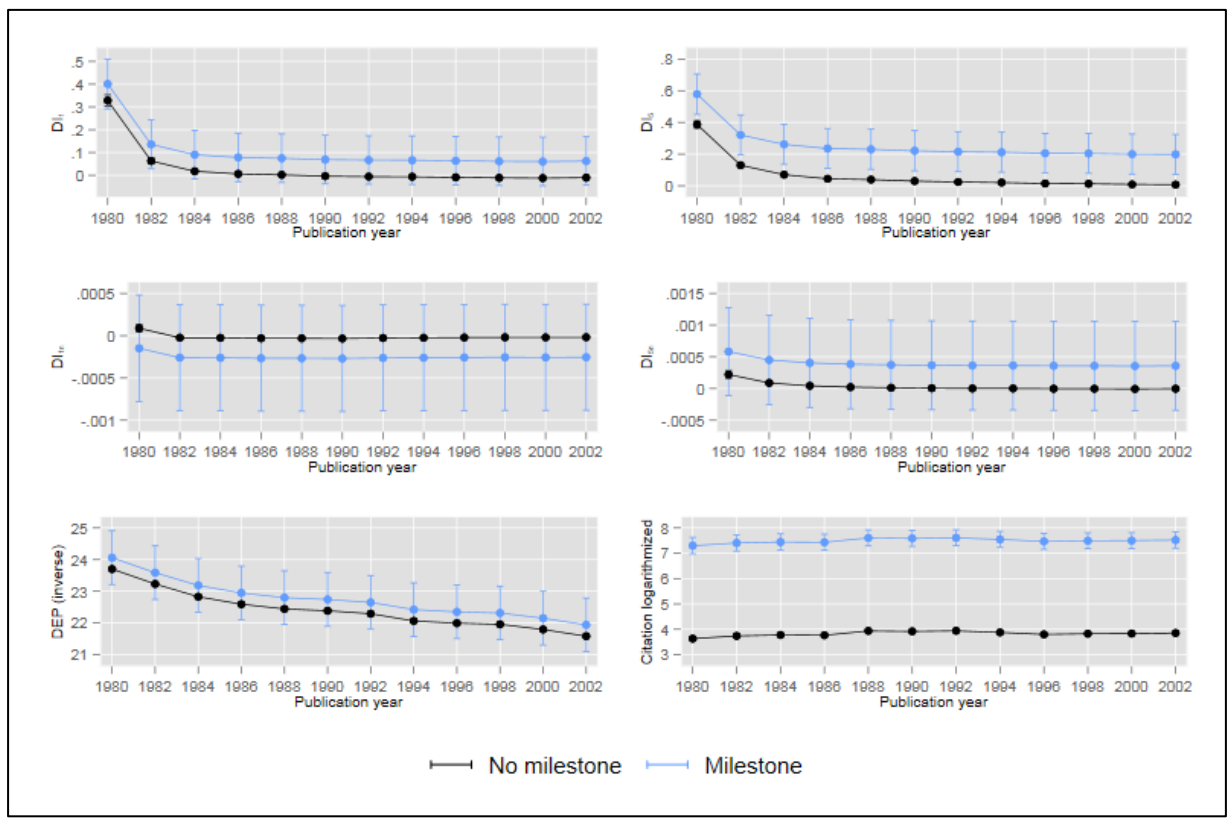

Figure 5. Averages of predicted performance values for milestone papers and other papers over publication years

In order to take a more detailed look at the relationship between milestone assignments and performance indicators, we calculated averages of predicted performance values for milestone papers and other papers over the publication years (see Figure 5). These predicted values result from the regression models in Table 2 (including control variables). In the interpretation of the results in Figure 5, the different scales of the indicators should be considered. The very high case numbers in the group of non-milestone papers led to very small confidence intervals that are scarcely seeable in the figure. Clear differences between milestone and non-milestone papers in the expected direction – indicated by missing overlaps of confidence intervals for (mean) values of non-milestone and milestone papers – are especially visible for citations and $DI_5$. $DI_{1n}$ reveals higher values for non-milestone than for milestone papers.

## 3.2    (Coarsened) exact matching (CEM)

Since the regression models are based on a very unbalanced dataset with only 39 papers in the milestone paper group and several thousands in the 'control group', we decided to perform additional analyses in which we reduced the control group to those papers that are really comparable to the milestone papers.

The basic idea of the CEM method is to find paper pairs in the data that (1) differ in the treatment (here: whether a paper is a milestone paper or not) but (2) are similar in certain characteristics (e.g. the number of co-authors). The method simulates a quasi-experimental design although only observational data are available. Pairs of papers are compared and these pairs differ in the treatment, but have a similar number of co-authors and pages, were published in a similar year and with a similar number of international collaborations, and come from similar regions (USA, China, and EU 28). In the analyses with the disruption index variants, citation counts have been additionally considered (see above). The CEM analysis is based on similar papers, since we calculated quintiles for the variables and used these quintiles as input for the CEM (see Caliendo & Kopeinig, 2008; Iacus et al., 2012).

Table 3. Results of (coarsened) exact matching (CEM)

| Variable | Matched (yes / no) | Average treatment effect (ATE) | Standard error | 95% confidence interval | $N$ |
|---|---|---|---|---|---|
| $DI_1$ | 38 / 1 | 0.14959* | 0.06192 | [0.02823, 0.27094] | 76 |
| $DI_5$ | 38 / 1 | 0.23884*** | 0.05506 | [0.13093, 0.34675] | 76 |
| $DI_{1n}$ | 38 / 1 | -0.00033 | 0.00024 | [-.00080, 0.00013] | 76 |
| $DI_{5n}$ | 38 / 1 | 0.00175** | 0.00053 | [0.00072, 0.00278] | 76 |
| DEP (inverse) | 38 / 1 | 0.51217*** | 0.12102 | [0.27497, 0.74937] | 76 |
| Citations logarithmized | 39 / 0 | 4.03215*** | 0.18535 | [3.66887, 4.39543] | 78 |

Note. * $p < 0.05$, ** $p < 0.01$, *** $p < 0.001$

The results of CEM for the six performance indicators are shown in Table 3. For the disruption index variants, 38 treated papers could be matched with an untreated paper. For citations, all papers could be matched with a corresponding paper in the control group. The pairs scarcely deviate in the considered variables (e.g. number of co-authors). The average treatment effect (ATE) is the estimated difference in the performance indicators between the treated and untreated paper groups. For example, the ATE for the logarithmized citations is 4.03. Thus, the average milestone effect in our dataset increases the logarithmized citations by around four. All ATEs in Table 3 are statistically significant (and positive), except the ATE for $DI_{1n}$ (which also has the only negative value). Taken as a whole, the results in Table 3 correspond well with the results of the regression models in Table 2: a negative effect for $DI_{1n}$ and relatively large effects for logarithmized citations and $DI_5$. The only difference from the previous results is a relatively large effect for DEP.

In section 2.2, we explained that it might be reasonable to use the milestone paper assignment as a dependent variable and the various performance indicators as one independent variable each. If we change the dependent and independent variables in this way, we assume that the performance indicators indicate the disruptive or impact potential and the milestone paper assignment is dependent on this potential. The results of the corresponding regression analyses and margin plots are given in the appendix (see Table A5 and Figure A6). We only included the performance indicators as independent variables in the regression models, since the other independent variables used in Table 2 might be related to performance indicators, but not to assignments by experts. Table A5 reports odds ratios which can be interpreted as follows: for each additional citation (logarithmized), the odds of a milestone paper assignment increase by a factor of around 10. As the predictive margins in Figure A6 reveal, substantial relationships between the performance indicators and assignments can be observed for $DI_5$, DEP (inverse), and logarithmized citations.

# 4 Discussion

In the ethos of science, Merton (1973) included a set of norms that would guide the behavior of scientists. According to Ziman (2000), "the norms of 'originality', 'scepticism' and 'communalism' are put into operation as processes of 'variation', 'selection' and 'retention' respectively" (p. 277). The norm of originality 'reminds' researchers that it is their superficial role to advance knowledge: "in the institution of science originality is at premium. For it is through originality, in greater or smaller increments, that knowledge advances" (Merton, 1957, p. 639). Since "originality energises the scientific enterprise" (Ziman, 1996, p. 68), researchers are especially rewarded for research who have advanced knowledge in great increments by publishing novel discoveries, broadly applicable methods, and ground-breaking theories (Wagner et al., 2019). Such great increments can be especially expected for research that disrupts previous research lines in a field. Whereas 'developmental' research is characterized by small increments, disruptive research advances by multiple-league steps. It is obvious that national science policies are especially interested in promoting disruptive research, and it is not at all surprising that indicators have been developed to identify such research activities – based on bibliometric data.

This study focuses on a recently introduced type of indicators measuring disruptiveness. In the current study, we included with $DI_1$ the initially proposed indicator of this new type (Wu et al., 2019) and several variants: $DI_5$, $DI_{1n}$, $DI_{5n}$, and DEP. Since indicators should measure what they propose to measure, we investigated the convergent validity of the indicators. We used a list with milestone papers that had been selected and published by editors of PRL, and investigated whether this human (expert)-based list is related to values of the several disruption indicator variants and, if so, which variants show the highest correlation with expert judgements. In recent years, two other studies based on Reference Publication Year Spectroscopy (Liao, Shen, & Yang, 2019) and time-balanced network centrality (Mariani, Medo, & Zhang, 2016) have already used the PRL list of milestone papers for the correlation of expert judgements with bibliometric results.

In this study, we used bivariate statistics, multiple regressions models, and (coarsened) exact matching to investigate the convergent validity of the indicators. The results show that the indicators correlate differently with the milestone paper assignments by the editors. It is not the initially proposed disruption index that performed best ($DI_1$), but the variant $DI_5$ which was introduced by Bornmann, Devarakonda, et al. (2019). In the CEM analysis of this study, the DEP variant – introduced by Bu et al. (2019) – also showed favorable results. The fact that these indicators perform well suggests that it is important to take into account how strong the citation links between citing papers and cited references are when assessing whether the citing papers depend on the focal paper's cited references. $DI_5$ and DEP both apply this strategy, while the other indicators do not. As $DI_5$ in particular performs well, it seems that only citing papers with a certain number of citation links to the focal paper's cited references are reliable indicators for a developmental focal paper.

The results of this study are in agreement with the results of the disruption indicator comparison of Bornmann, Devarakonda, et al. (2019) showing similar favorable results for $DI_5$. Our results also accord with the results by Chunli, Zhenyue, Dongbo, and Jiang (2020) who investigated $DI_1$ by comparing $DI_1$ values of papers leading to Nobel prizes and a control group of comparable papers published in the same journal. Their results show "no evidence that the 557 prize winning papers are more disruptive, because the disruption of the prize winning group averages 0.236 while that of the control group averages 0.239 and two-tail paired samples t-test showed p=0.89" (Chunli et al., 2020).

What are the limitations of this study? (1) One should keep in mind especially if a concept such as the disruptiveness of research is measured, indicators may be biased or imperfect: "the awareness that all performance indicators are 'partial' or 'imperfect' … is as old as the use of performance indicators itself. Indicators may be imperfect or biased, but in the application of such indicators this is not seldom forgotten" (Moed, 2017, p. 6). However, one should also keep in mind that in order to measure the convergent validity in this study, we used judgements by experts who may be also concerned by biases (see Bornmann & Daniel, 2009). (2) The second limitation is related to the first. The definition of disruptiveness, originality, excellence, and similar other concepts in science is a normative decision (and is thus subjective); it follows that the attribution of these concepts it contingent. For Ferretti, Pereira, Vértesy, and Hardeman (2018), for instance, "quantifying research excellence is first and foremost a political and normative issue" (p. 739).

(3) The third limitation concerns the used data for calculating the disruption indicators. Cited references do not reflect all possible influences that lead to the research reported in a paper. According to Schilling and Green (2011) "it is also possible for individuals to draw from deeper or broader knowledge reservoirs than what is reflected in the paper's references, which introduces some noise" (p. 1325). This limitation of cited references could also be demonstrated by Tahamtan and Bornmann (2018b) who interviewed authors of landmark papers in scientometrics. Tahamtan and Bornmann (2018b) were interested in the question as to whether the roots of the landmark papers are reflected in their cited references. (4) The fourth limitation is related to the definition of milestone papers by the PRL editors. The assignments by the PRL editors do not explicitly aim at identifying disruptive papers, but milestone papers. It can be expected that this assignment is closely related to the notion of disruptiveness. However, the assignment as a milestone paper is only a proxy for disruptiveness, and we do not know exactly how other characteristics of the papers have influenced the assignment.

With the introduction of the disruption indicators in scientometrics, new types of indicators have been proposed that focus on a concept of enormous importance in science policy: the identification of research opening up new directions in fields. Before these indicators are used in the practice of research evaluation, their validity should be ensured. Thus, we encourage the realization of other studies investigating the validity of the indicators using datasets from fields other than physics, and other time periods and databases than those used in this study.

## Acknowledgements

The bibliometric data used in this paper are from an in-house database developed and maintained in cooperation with the Max Planck Digital Library (MPDL, Munich) and derived from the Science Citation Index Expanded (SCI-E), Social Sciences Citation Index (SSCI), Arts and Humanities Citation Index (AHCI) prepared by Clarivate Analytics, formerly the IP & Science business of Thomson Reuters (Philadelphia, Pennsylvania, USA). We thank Henry Small for providing feedback on earlier versions of this paper.

# Appendix

Table A4. Predictors of disruption index variants and logarithmized citations (excluding two extraordinary cases)

| Variable | $DI_1$ | $DI_1$ | $DI_5$ | $DI_5$ | $DI_{1n}$ | $DI_{1n}$ | $DI_{5n}$ | $DI_{5n}$ | DEP (inverse) | DEP (inverse) | Citations logarithmized | Citations logarithmized |
|---|---|---|---|---|---|---|---|---|---|---|---|---|
| Milestone letter (=1) | 0.09493* | 0.08316* | 0.25156*** | 0.18970*** | -0.00038 | -0.00033 | 0.00134*** | 0.00122*** | 1.05285*** | 0.23786** | 3.73375*** | 3.59756*** |
|  | (0.04) | (0.04) | (0.04) | (0.03) | (0.00) | (0.00) | (0.00) | (0.00) | (0.11) | (0.08) | (0.17) | (0.18) |
| SD | 0.12485 | 0.12485 | 0.13151 | 0.13151 | 0.00015 | 0.00015 | 0.00026 | 0.00026 | 1.36915 | 1.36915 | 1.1566 | 1.1566 |
| Number of years |  | 0.00518*** |  | 0.00705*** |  | 0.00000* |  | 0.00000*** |  | 0.07156*** |  | -0.00424*** |
|  |  | (0.00) |  | (0.00) |  | (0.00) |  | (0.00) |  | (0.00) |  | (0.00) |
| Citations logarithmized |  | 0.00090 |  | 0.01436*** |  | -0.00001*** |  | 0.00003*** |  | 0.20797*** |  |  |
|  |  | (0.00) |  | (0.00) |  | (0.00) |  | (0.00) |  | (0.01) |  |  |
| Number of co-authors |  | 0.00001 |  | 0.00005*** |  | 0.00000 |  | 0.00000 |  | 0.00173*** |  | -0.00298*** |
|  |  | (0.00) |  | (0.00) |  | (0.00) |  | (0.00) |  | (0.00) |  | (0.00) |
| Number of pages |  | 0.00413*** |  | -0.00681*** |  | 0.00000* |  | -0.00001*** |  | -0.30724*** |  | 0.52315*** |
|  |  | (0.00) |  | (0.00) |  | (0.00) |  | (0.00) |  | (0.02) |  | (0.02) |
| Number of countries |  | -0.00196** |  | -0.00154* |  | -0.00000* |  | -0.00000 |  | -0.03292*** |  | -0.02622*** |
|  |  | (0.00) |  | (0.00) |  | (0.00) |  | (0.00) |  | (0.01) |  | (0.01) |
| USA (=1) |  | -0.00343** |  | -0.00426** |  | -0.00000** |  | -0.00000* |  | -0.01132 |  | 0.23020*** |
|  |  | (0.00) |  | (0.00) |  | (0.00) |  | (0.00) |  | (0.02) |  | (0.01) |
| China (=1) |  | 0.00568** |  | 0.00272 |  | 0.00000 |  | 0.00001** |  | -0.02516 |  | -0.05749 |
|  |  | (0.00) |  | (0.00) |  | (0.00) |  | (0.00) |  | (0.06) |  | (0.05) |
| EU 28 (=1) |  | -0.00098 |  | -0.00012 |  | 0.00000 |  | -0.00000 |  | -0.00194 |  | 0.15450*** |
|  |  | (0.00) |  | (0.00) |  | (0.00) |  | (0.00) |  | (0.02) |  | (0.01) |
| Constant | 0.00858*** | -0.05527*** | 0.03934*** | -0.05252*** | -0.00002*** | 0.00002* | 0.00001*** | -0.00010*** | 22.20828*** | 21.97510*** | 3.84082*** | 1.71602*** |
|  | (0.00) | (0.00) | (0.00) | (0.00) | (0.00) | (0.00) | (0.00) | (0.00) | (0.01) | (0.06) | (0.01) | (0.09) |
| $R^2$ | 0.00045 | 0.06895 | 0.00286 | 0.12935 | 0.00503 | 0.01512 | 0.02090 | 0.05242 | 0.00047 | 0.13864 | 0.00813 | 0.08181 |
| $N$ | 44804 | 44804 | 44804 | 44804 | 44806 | 44806 | 44806 | 44806 | 43844 | 43844 | 44806 | 44806 |

Notes. Robust standard errors in parentheses
* $p < 0.05$, ** $p < 0.01$, *** $p < 0.001$

Table A5. Milestone paper prediction based on various disruption index variants and logarithmized citations (the table reports the results of six models)

| Independent variable | Milestone paper (dependent variable) | Pseudo $R^2$ | N |
|---|---|---|---|
| $DI_1$ | 13.451*** | .028 | 44,809 |
|  | (6.272) |  |  |
| $DI_5$ | 40.621*** | .092 | 44,809 |
|  | (9.574) |  |  |
| $DI_{1n}$ | 0.000 | .001 | 44,812 |
|  | (0.000) |  |  |
| $DI_{5n}$ | 0.000 | .047 | 44,812 |
|  | (0.000) |  |  |
| DEP (inverse) | 3.880*** | .071 | 43,849 |
|  | (1.003) |  |  |
| Citations logarithmized | 10.240*** | .539 | 44,812 |
|  | (1.878) |  |  |

Notes. Odds ratios, robust standard errors in parentheses
* $p < 0.05$, ** $p < 0.01$, *** $p < 0.001$

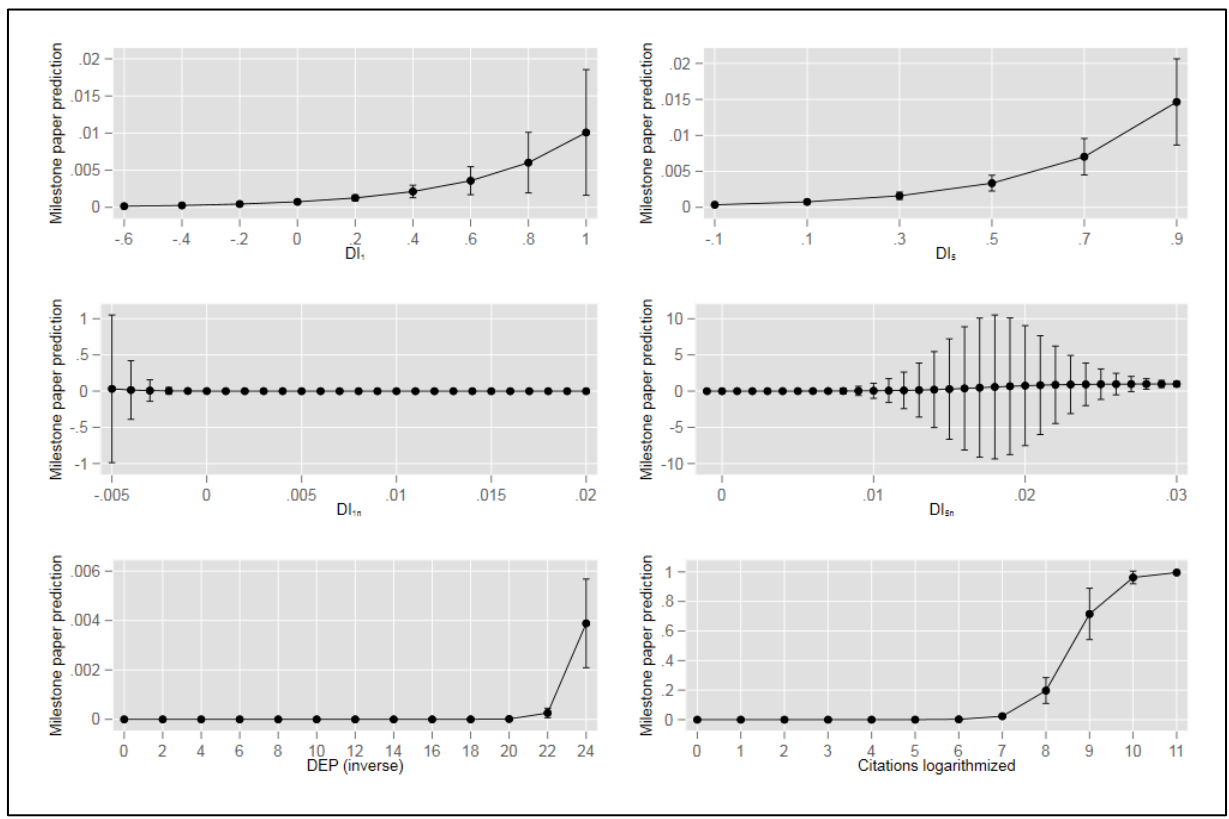

Figure A6. Milestone paper prediction based on various disruption index variants and logarithmized citations

Merton, R. K. (1973). *The sociology of science: Theoretical and empirical investigations*. Chicago, IL, USA: University of Chicago Press.

Mitchell, M. N. (2012). *Interpreting and visualizing regression models using Stata*. College Station, TX, USA: Stata Corporation.

Moed, H. F. (2017). *Applied evaluative informetrics*. Heidelberg, Germany: Springer.

Mutz, R. (2016). Do we really need BIBLIO-metrics to evaluate individual researchers? *Infozine*(1).

Onodera, N. (2016). Properties of an index of citation durability of an article. *Journal of Informetrics, 10*(4), 981-1004. doi: 10.1016/j.joi.2016.07.001.

Onodera, N., & Yoshikane, F. (2014). Factors affecting citation rates of research articles. *Journal of the Association for Information Science and Technology, 66*(4), 739–764. doi: 10.1002/asi.23209.

Panel for Review of Best Practices in Assessment of Research, Panel for Review of Best Practices in Assessment of Research, Development Organizations, Laboratory Assessments Board, Division on Engineering, Physical Sciences, & National Research Council. (2012). *Best Practices in Assessment of Research and Development Organizations*: The National Academies Press.

Peters, H. P. F., & van Raan, A. F. J. (1994). On determinants of citation scores - a case study in chemical engineering. *Journal of the American Society for Information Science, 45*(1), 39-49.

Puccio, G. J., Mance, M., & Zacko-Smith, J. (2013). Creative leadership. Its meaning and value for science, technology, and innovation. In S. Hemlin, C. M. Allwood, B. Martin & M. D. Mumford (Eds.), *Creativity and leadership in science, technology, and innovation* (pp. 287-315). New York, NY, USA: Taylor & Francis.

Rowlands, I. (2018). What are we measuring? Refocusing on some fundamentals in the age of desktop bibliometrics. *FEMS Microbiology Letters, 365*(8). doi: 10.1093/femsle/fny059.

Ruscio, J., Seaman, F., D'Oriano, C., Stremlo, E., & Mahalchik, K. (2012). Measuring scholarly impact using modern citation-based indices. *Measurement: Interdisciplinary Research and Perspectives, 10*(3), 123-146.

Schilling, M. A., & Green, E. (2011). Recombinant search and breakthrough idea generation: An analysis of high impact papers in the social sciences. *Research Policy, 40*(10), 1321-1331. doi: 10.1016/j.respol.2011.06.009.

Schneider, J. W., & Costas, R. (2016). Identifying potential "breakthrough" publications using refined citation analyses: Three related explorative approaches. *Journal of the Association for Information Science and Technology, 68*(3), 709-723.

Seglen, P. O. (1992). The skewness of science. *Journal of the American Society for Information Science, 43*(9), 628-638.

Shadish, W. R., Cook, T. D., & Campbell, D. T. (2002). *Experimental and quasi-experimental designs for generalized causal inference*. Boston, MA, USA: Houghton Mifflin Company.

Sheskin, D. (2007). *Handbook of parametric and nonparametric statistical procedures* (4th ed.). Boca Raton, FL, USA: Chapman & Hall/CRC.

Stanek, K. Z. (2008). How long should an astronomical paper be to increase its impact? Retrieved September 22, 2008, from http://arxiv.org/abs/0809.0692

StataCorp. (2017). *Stata statistical software: release 15*. College Station, TX, USA: Stata Corporation.

Tahamtan, I., & Bornmann, L. (2018a). Core elements in the process of citing publications: Conceptual overview of the literature. *Journal of Informetrics, 12*(1), 203-216. doi: 10.1016/j.joi.2018.01.002.35